\documentclass[12pt]{article}
\usepackage[dvips]{graphicx}
\usepackage{pdproc}

  \makeatletter 
  \def\@cite#1{[#1]} 
  \makeatother    
\makeatletter
\newcommand{\vereq}[2]{\lower3pt\vbox{\baselineskip1.5pt \lineskip1.5pt
\ialign{$\m@th#1\hfill##\hfil$\crcr#2\crcr\sim\crcr}}}
\newcommand{\lesssim}{\mathrel{\mathpalette\vereq<}}

\makeatother

\begin{document}

\title{
\hfill{\normalsize\vbox{\hbox{\rm DPNU-04-16}  }}\\
\vspace{0.1cm}
A dynamical Origin of Little Higgs \\
- Hidden Local Symmetry in Large $N_f$ QCD -\footnote{%
Talk given at 
``The 12th International Conference
on Supersymmetry
and Unification of Fundamental Interactions'',
June 17-23, 2004,
Epochal Tsukuba, Tsukuba, Japan.
}
}

\author{Masayasu Harada}

\address{ 
Department of Physics, Nagoya University,
Nagoya, 464-8602, JAPAN\\
{\rm E-mail: harada@eken.phys.nagoya-u.ac.jp}}

\abstract{
In this write-up I summarize the main points of our recent work,
in which we showed that
the large flavor QCD is regarded as 
a dynamical origin of little Higgs,
and that the mass of such a little Higgs 
becomes actually very small.
}

\normalsize\baselineskip=15pt

\section{Introduction}

In little Higgs 
models~\cite{Little-higgs}, 
the Higgs is introduced as pseudo
Nambu-Goldstone (NG) boson,
which naturally explains the lightness of Higgs boson.

In Ref.~\cite{HTY}, we pointed out that it is easy to 
formulate little Higgs models using the hidden local
symmetry (HLS).
We picked up 
a little Higgs model with two sites
and two links, where
one site corresponds to the U(1) gauge symmetry 
and another to the SU($N_f$) gauge symmetry.
When the U(1) gauge symmetry is a
subgroup of the chiral SU($N_f$)$_L \times$SU($N_f$)$_R$ symmetry,
the above little higgs model is equivalent to
the HLS with 
the parameter choice $a = 1$ taken.
Based on this,
we showed that the large flavor QCD is regarded as 
a dynamical origin of little Higgs,
and that the mass of such a little Higgs 
becomes actually very small near the chiral symmetry restoration
point.
In this write-up, I will show main points of our work.

\section{Hidden Local Symmetry}
\label{sec:HLS}

The HLS model~\cite{BKUYY-BKY} 
is based on the
$G_{\rm global} \times H_{\rm local}$ symmetry, where
$G = \mbox{SU($N_f$)}_{\rm L} \times 
\mbox{SU($N_f$)}_{\rm R}$  is the 
global chiral symmetry and 
$H = \mbox{SU($N_f$)}_{\rm V}$  the HLS whose gauge bosons are
identified with the vector mesons (the $\rho$ meson and its flavor
partners) denoted as $V_\mu$.
Here $N_f$ denotes the number of massless quark flavors in the
underlying 
QCD.
The basic dynamical variables in the HLS model are gauge bosons
$\rho_\mu=\rho_\mu^a T_a$  
of the HLS and two 
SU($N_f$)-matrix-valued variables $\xi_{\rm L}$ and 
$\xi_{\rm R}$.
These can be parameterized as
$\xi_{\rm L,R} = e^{i\sigma/F_\sigma} e^{\mp i\pi/F_\pi}$
where $\pi = \pi^a T_a$
denotes the NG bosons ($\pi$ meson and its flavor 
partners) associated with 
the spontaneous breaking of $G$ and 
$\sigma = \sigma^a T_a$ (with $J^{PC} =0^{+-}$)
the NG bosons absorbed into the (longitudinal) HLS gauge bosons.
$F_\pi$ and $F_\sigma$ are the relevant decay constants, with a ratio
$a$ defined by $a \equiv F_\sigma^2/F_\pi^2$.

The Lagrangian of the HLS is expressed as
\begin{equation}
{\mathcal L} =
  \frac{a+1}{4} \, F_\pi^2 \, \mbox{tr}
    \left[ \hat{\alpha}_{{\rm L}\mu} \hat{\alpha}_{\rm L}^\mu 
    + \hat{\alpha}_{{\rm R}\mu} \hat{\alpha}_{\rm R}^\mu \right]
  + \frac{a-1}{2} \, F_\pi^2 \, \mbox{tr}
    \left[ \hat{\alpha}_{{\rm L}\mu} \hat{\alpha}_{\rm R}^\mu \right]
  - \frac{1}{2} \mbox{tr} 
    \left[ V_{\mu\nu} V^{\mu\nu} \right]
\ ,
\end{equation}
where the 1-forms $\hat{\alpha}_{\rm L,R}^\mu$ are
defined as
$\hat{\alpha}_{\rm L,R}^\mu = - i D^\mu \xi_{\rm L,R} \cdot
  \xi^\dag_{\rm L,R}$
with the covariant derivatives given by
$D^\mu \xi_{\rm L,R} = (\partial^\mu - i g V_\mu)\xi_{\rm L,R}$.
I would like to stress that,
when we take $a=1$ in the above Lagrangian, the second 
term disappears,
and the fields in $\xi_{\rm L}$ couples to those
in $\xi_{\rm R}$ only through the HLS gauge boson $V_\mu$,
which is nothing but the theory space locality.
It should be noticed that the parameter choice $a=1$ together
with $g=0$ corresponds to the Georgi's ``vector limit''~\cite{Georgi}.

\section{Stability of theory space locality}
\label{sec:STSL}

When we consider quantum corrections,
it is essential to have a systematic expansion.
Fortunately, a systematic chiral perturbation can be done
in the HLS including the dynamical 
effect of the HLS gauge boson~\cite{Georgi,HY:92,Tanabashi,HY:PRep}.
This chiral perturbation is actually justified in the large $N_c$
limit of QCD, where
the pion decay constant $F_\pi$ becomes large while the vector meson
mass $m_V$ is kept fixed.
As a result, the ratio $m_V/4\pi F_\pi$
becomes small in the large $N_c$ QCD.
In other words, the HLS gauge coupling becomes small in the
large $N_c$ QCD.

One may think that 
the scalar meson should be included since it
is lighter than the vector meson in real-life QCD.
However, recently in Ref.~\cite{HSS:04},
we extended the analysis adopted in Refs.~\cite{SS:95,HSS:96}
for studying the $\pi$-$\pi$ scattering in the real-life QCD
to the one in the large $N_c$ QCD, and 
showed that, for $N_c \ge 6$,
the unitarity in the scalar channel of the $\pi$-$\pi$ scattering 
is satisfied without
scalar mesons up until the energy scale of $4\pi F_\pi$.
This indicates that we do not need the scalar meson in the
low-energy region in the large $N_c$ QCD.
In the real-life QCD, we know that there is a light
scalar meson ($\sigma$ meson), which is actually very
broad.
I expect that loop corrections from 
such broad resonances are very small, and
that the chiral perturbation in the HLS
is still possible, as far as we do not see the
scalar channel.

I should note that there is no guarantee for the
smallness of $(a-1)$ in the large $N_c$ QCD.
In other words, the theory space locality may not be justified in
the large $N_c$ QCD.
However, 
since $a=1$ is a fixed point of the renormalization
group equation for the parameter $a$~\cite{HY:PRep},
there is no divergent correction
to $a=1$ at one-loop level.
In Ref.~\cite{HS:VD}, on the other hand, 
we have shown that there exists
a small finite correction when we take non-zero gauge coupling.
These results imply that,
once the parameter $a$ is tuned to be 1,
there is only small violation of 
the theory space locality even at two-loop level,
i.e.,
the theory space locality is quite stable.

\section{Little Higgs in Large $N_f$ QCD}
\label{sec:LHLNQ}

It is well-known that, in the large flavor QCD,
the chiral symmetry restoration occurs at some critical number
of flavors $N_f^{\rm crit}$~\cite{largeNf}.
In Ref.~\cite{HY:VM}, we showed that,
due to the vector manifestation, the parameters of the 
HLS $g$ and $a$ approach the fixed point values of the
RGEs near the critical point, i.e., $(g,a) \rightarrow (0,1)$
for $N_f \rightarrow N_f^{\rm crit}$.

Since the parameter $a$ approaches $1$ near the critical point,
the theory space locality is actually guaranteed near the
restoration point.
This implies that the large flavor QCD
can be regarded as an ultraviolet (UV) 
completion of a little Higgs model.
In Ref.~\cite{HTY},
we calculated the mass of the little Higgs in the large flavor
QCD by gauging the U(1) subgroup of the chiral symmetry.
The generator of the U(1) gauge symmetry and the
generator corresponding to the little Higgs 
(pseudo NG boson)
are given by
\begin{equation}
Q = \left(\begin{array}{ccc}
  2/3 & & \\ & -1/3 & \\ & & \ddots \\
\end{array}\right)
\ ,
\quad
T_{\rm PNG} = \left(\begin{array}{cccc}
  0 & 1 & 0 & \cdots \\
  0 & 0 & 0 & \\
  0 & 0 & 0 & \\
  \vdots & & & \ddots \\
\end{array}\right)
\ .
\label{generators}
\end{equation}
It should be noticed that,
when we set the parameter $a$ to $1$, the HLS
theory becomes a little Higgs model with two sites
and two links.

The loop correction to the mass of the little Higgs in the HLS
is given by
\begin{equation}
\left. m_H^2 \right\vert_{\rm HLS} =
  \frac{ \alpha_{\rm U(1)} }{4\pi} 
  \left[
    (1-a) \Lambda^2 + 3 a M_\rho^2 \ln \Lambda^2 + \cdots
  \right]
\ .
\label{HLS cont 1}
\end{equation}
Here, I would like to stress that the quadratic divergence
disappears when we take $a=1$, due to the theory space
locality, and the mass is expressed as
\begin{equation}
\left. m_H^2 \right\vert_{\rm HLS} =
 \alpha_{\rm U(1)} \, \alpha_{\rm HLS} \, 3 F_\pi^2 \,\ln \Lambda^2
\ .
\label{HLS cont 2}
\end{equation}
In several little Higgs models, this is the end of story.
In the present case, however, I consider that the little
Higgs model is a low energy effective field theory 
derived from the large flavor QCD.
As a result, there is a remnant of the underlying theory,
which gives a contribution to the mass of the little Higgs.
In our work, we calculated the UV contribution
by using the current algebra formula~\cite{Das,Yamawaki:82}:
\begin{equation}
\left. m_H^2 \right\vert_{\rm UV} =
  \frac{ 3 \alpha_{\rm U(1)} }{4\pi F_\pi^2} 
  \int^{\infty}_{\Lambda} d Q^2 \,
  \left[ \Pi_V(Q^2) - \Pi_A(Q^2) \right]
=
  \alpha_{\rm U(1)} \alpha_{\rm QCD} \, 
  \frac{3(N_c^2-1)}{N_c^2} \,
  \frac{\langle \bar{q}q\rangle^2_\Lambda}{ F_\pi^2\,\Lambda^2}
\ ,
\label{UV cont}
\end{equation}
where $\Pi_V$ and $\Pi_A$ are the vector and 
axial vector current correlators.
The second expression is obtained by substituting
the current correlators derived in the
operator product expansion.
The mass of the little Higgs is given by the sum of
the UV contribution in Eq.~(\ref{UV cont})
and the low-energy contribution in Eq.~(\ref{HLS cont 1}).

The mass of the little Higgs (pseudo NG boson)
in the setup given in Eq.~(\ref{generators}) 
becomes the
$\pi^+$-$\pi^0$ mass difference in the real-life QCD,
which is estimated as
\begin{equation}
\Delta m_\pi^2 = \left. m_H^2 \right\vert_{\rm HLS}
  + \left. m_H^2 \right\vert_{\rm UV}
  \simeq 1000 \,\mbox{MeV}^2 + 200\,\mbox{MeV}^2
  = 1200 \,\mbox{MeV}^2
\ .
\end{equation}
The predicted value is 
very close to the experimental value, 
$\left. \Delta m_\pi^2 \right\vert_{\rm exp} = 1261 \,\mbox{MeV}^2$,
which shows the validity of our calculation.

For making the estimation of
the mass of the little Higgs
in the large flavor QCD,
we used the following scaling properties of the parameters
in the large flavor QCD:
\begin{equation}
F_\pi \sim m_{\rm dyn} \ ,
\quad
M_\rho \sim g F_\pi \sim \langle \bar{q} q \rangle F_\pi
\sim m_{\rm dyn}^{4-\gamma_m}
\ ,
\quad
(a-1) \sim \langle \bar{q}q\rangle \sim m_{\rm dyn}^{6-2\gamma_m}
\ ,
\end{equation}
where $m_{\rm dyn}$ is the dynamical fermion mass.
$\gamma_m$ is the anomalous dimension of the
fermion mass operator, which satisfies
$\gamma_m \lesssim 1 $ in the large flavor QCD.
Substituting these scaling properties into the sum of the
contributions in Eqs.~(\ref{UV cont}) and (\ref{HLS cont 1}),
one can easily show that the UV contribution
becomes dominant near the critical point.
Then, the mass of little Higgs scales as
$m_H^2 \sim m_{\rm dyn}^{4-2\gamma_m}$.
As a result, the scaling property of 
the ratio of the little Higgs mass to the new physics scale
is given by 
\begin{equation}
\frac{m_H}{4\pi F_\pi} \sim m_{\rm dyn}^{1-\gamma_m}
\ \rightarrow\ 
\left\{\begin{array}{ll}
  0 & \mbox{for}\ \gamma_m < 1 \ ,\\
  0.01 \ \ & \mbox{for} \ \gamma_m = 1 \ ,
\end{array}\right.
\end{equation}
where the estimation for $\gamma_m = 1$ is done numerically.
This implies that
the large hierarchy between the mass of little
Higgs and that the scale of new physics can be easily achieved
in the large flavor QCD.

\section{Acknowledgments}

I would like to thank the organizers for giving me an 
opportunity to present my talk.
I am grateful to Masaharu Tanabashi and Koichi Yamawaki
for collaboration in Ref.~\cite{HTY} on which this talk is
based.
My work is supported in part by
the JSPS Grant-in-Aid for Scientific Research (c) (2) 16540241,
and by 
the 21st Century COE
Program of Nagoya University provided by Japan Society for the
Promotion of Science (15COEG01).

\bibliographystyle{plain}

\end{document}